%% file: paper.tex
\documentclass[showpacs,preprint,amsmath,amssymb,aps,superscriptaddress,prd]{revtex4}

\usepackage{graphicx}% Include figure files
\usepackage{dcolumn}% Align table columns on decimal point
\usepackage{bm}% bold math

\begin{document}
\newcommand{\ET}{$E_T$\ }
\newcommand{\PT}{$P_T$\ }
\newcommand{\ETns}{$E_{T}$}
\newcommand{\PTns}{$P_{T}$}
\newcommand{\mgev}{GeV/$c^2$}
\newcommand{\mgevs}{GeV/$c^2$\ }
\newcommand{\met}{\mbox{$\raisebox{.3ex}{$\not$}E_{T}$\ }}
\newcommand{\metns}{\mbox{$\raisebox{.3ex}{$\not$}E_{T}$}}

%\preprint{APS/123-QED}

\title{
\bf Measurement of the ${\boldmath t\bar{t}}$ Production Cross Section
in ${\boldmath p\bar{p}}$ Collisions at ${\boldmath \sqrt{s} = 1.96}$ 
TeV Using Kinematic Fitting of $b$-tagged Lepton+Jet Events }
% Force line breaks with \\

\input{author.tex}

\date{\today}% It is always \today, today,

\begin{abstract}
We report a measurement of the $t \bar{t}$ production cross section
using the CDF II detector at the Fermilab Tevatron.  The data consist
of events with an energetic electron or muon, missing transverse
energy, and three or more hadronic jets, at least one of which is
identified as a $b$-quark jet by reconstructing a secondary vertex.  The
background fraction is determined from a fit of the transverse energy
of the leading jet.  Using 162$\pm$10 pb$^{-1}$ of data, 
the total cross section is found to be
6.0$\pm$1.6(stat.)$\pm$1.2(syst.) pb, which is consistent with the 
Standard Model prediction. 
%An article usually includes an abstract, a concise summary of the work
%covered at length in the main body of the article. It is used for
%secondary publications and for information retrieval purposes. Valid
%PACS numbers may be entered using the \verb+\pacs{#1}+ command.
\end{abstract}

\maketitle

\section{Introduction}
The top quark is the most massive of nature's building blocks yet
discovered.  Because new physics associated with electroweak symmetry
breaking will likely couple to an elementary particle in proportion to
its mass, it is important to measure the top quark couplings as
accurately as possible.  In the strong interaction sector, the
couplings are reflected in the $t\bar{t}$ production cross section in
hadron collisions.  Previous measurements were made in $p\bar{p}$
collisions at a center-of-mass energy of 1.8
TeV~\cite{CITE:cdf_d0_runI}.  We recently reported a 
result~\cite{CITE:dilepton}, using
data taken at 1.96 TeV with the CDF II detector at the Tevatron
collider, using the double leptonic decay mode of the top quark. 
Here we report a measurement of the $t\bar{t}$ production cross 
section using a different decay mode and a new method.

In order to measure the cross section, one first has to obtain a sample rich
in top quarks and then determine the amount of background in the
sample.  We select events consistent with the decay
chain $t\bar{t}\rightarrow WbW\bar{b}\rightarrow l\nu
bq\bar{q}^{\prime}\bar{b}$, where the charged lepton $l$ is either an
electron or muon.  We start with events containing an energetic
electron or muon, significant transverse momentum imbalance indicative
of a non-interacting neutrino, 
and at least three hadronic jets.  To
enrich the sample in top quarks, we require that at least one jet
contain a secondary vertex consistent with the decay of a
$B$ hadron.

Measurements in the past have relied on the ability of theoretical
calculations to determine, for the background, the
fraction of events that contain $b$-quark jets.  In this paper, we
instead measure the background fraction directly in the signal data
sample.  The transverse energy of the highest \ET jet~\cite{CITE:plug}
or the second highest \ET jet is a good discriminator between signal
and background.  
Typically in a $t\bar{t}$ event, these jets are the
primary decay products ($b$-jets) of the very heavy top quarks and
thus have a hard \ET spectrum.  For most of the background sources,
however, they are produced as QCD radiation, resulting in a much
softer bremsstrahlung-like \ET distribution.  We use the leading jet
\ET spectrum for the primary measurement of the signal fraction.  The
second leading jet distribution is used to check the result.

In order to obtain the background spectrum, we need data that are
kinematically similar to our final sample, but which do not have
significant $t\bar{t}$ contamination.  We show that the leading jet
\ET spectra for the background processes are similar whether or not the events
contain $b$-quark jets.  Then the non-heavy flavor spectrum becomes the
background template for measuring the $t\bar{t}$ fraction in the
signal sample.

We use the HERWIG~\cite{CITE:HERWIG} and PYTHIA~\cite{CITE:PYTHIA}
Monte Carlo calculations followed by a simulation of the CDF II
detector to obtain the $t \bar{t}$ signal behavior. The soft \ET
spectrum of the parton showers in these Monte Carlo models is not
relevant for the signal shape. To test that our method is plausible,
we study the background shape using the ALPGEN+HERWIG Monte Carlo
~\cite{CITE:ALPGEN}. ALPGEN provides a harder and more realistic jet
\ET spectrum.  For the study of $b$-jet identification, the PYTHIA
calculation is used.

\section{Detector}
The CDF II detector~\cite{CITE:CDFdetector} is an azimuthally and
forward-backward symmetric apparatus designed to study $p \bar{p}$
collisions at the Fermilab Tevatron.  It consists of a magnetic
spectrometer surrounded by calorimeters and muon chambers.  The
charged particle tracking system is immersed in a 1.4 T magnetic field
parallel to the $p$ and $\bar{p}$ beams. A 700,000-channel
silicon microstrip detector (SVX+ISL) provides tracking over the radial
range from 1.5 to 28 cm.  A 3.1 m long open-cell drift chamber, the Central
Outer Tracker (COT), covers the radial range from 40 to 137 cm. The
COT provides up to 96 measurements of the track position
with alternating axial and 2$^{\circ}$
stereo superlayers of 12 wires each. The fiducial region of the
silicon detector extends to $|\eta| \sim 2$, while the COT provides
coverage for $|\eta| \le 1$.
  
Segmented electromagnetic and hadronic calorimeters surround the
tracking system and measure the energy of interacting particles. 
The electromagnetic and hadronic calorimeters are
lead-scintillator and iron-scintillator sampling devices,
respectively, covering the pseudorapidity range $|\eta| < 3.6$. The
electromagnetic calorimeters are instrumented with proportional and
scintillating strip detectors that measure the transverse profile of
electromagnetic shower candidates at a depth corresponding to the shower
maximum. Drift chambers located outside the central hadron
calorimeters and behind a 60 cm iron shield detect muons
with $|\eta| < 0.6$. Additional drift chambers and
scintillation counters detect muons in the region 0.6$ < |\eta| <
1.0$. Gas Cherenkov counters measure the average number of inelastic
$p \bar{p}$ collisions and thereby determine the 
luminosity with the coverage 3.7$< |\eta| <$ 4.7.

The results reported here are based on data taken in Fermilab Collider
Run II between March, 2002 and September, 2003.  The integrated
luminosity is 162 pb$^{-1}$ for events selected with an electron or
central muon. For muon events with $|\eta|$ between 0.6 and 1.0, the
integrated luminosity is 150 pb$^{-1}$.

\section{Event Selection}
\subsection{Lepton trigger}
CDF employs a three level trigger system, the first two consisting of
special purpose hardware and the third a farm of computers.  For the
electron top sample, the level-1 trigger requires a track of $P_T>8$
GeV/$c\ $ matched to an electromagnetic calorimeter cell containing
$E_T>8$ GeV with a small amount of energy in the hadronic cell behind
it.  Calorimeter energy clustering is done at level-2, and the $\ge$ 8
GeV/$c\ $ track must be matched to an electromagnetic cluster with \ET
above 16 GeV.  At level-3, a reconstructed
electron candidate with $E_T>18$ GeV is required.  The $P_T>18$
GeV/$c\ $ level-3 muon triggers come directly from two level-1 triggers:
a track with $P_T>4$ GeV/$c\ $ is matched to a stub in the
central muon chambers; or a track with $P_T>8$ GeV/$c\ $ is matched to
a stub in the $0.6<|\eta|<1.0$ muon chambers.  

\subsection{W + jets selection}
After full event reconstruction, we require lepton candidates to pass
identification criteria and to be isolated from other energy deposits
in the calorimeter. The event selection
criteria are the same as those in Ref.~\cite{CITE:kinematic}, where they
are described in detail. Electron candidates must have a well-measured
track pointing at a cluster of energy in the calorimeter with \ET $>$
20 GeV. The lateral and transverse shower size in the calorimeters as
well as the transverse profile in the shower-maximum detectors must be
consistent with an electromagnetic cascade.  Muon candidates with \PT
$>$ 20 GeV/$c\ $ must pass through calorimeter cells whose energy
deposition is consistent with the ionization of a muon, and the
reconstructed position of the track segment in the muon chambers is
required to be consistent with multiple Coulomb scattering of the
extrapolated track from the COT.  Lepton candidates must also be
isolated.  Isolation ($I$) is defined as the ratio between calorimeter
energy in a cone of radius 0.4 in the $\eta$-$\phi$ plane around the lepton, 
but excluding
the lepton, divided by the lepton energy.  We require $I\le0.1$.  In
addition, all candidate events 
must have \met $>$ 20 GeV. The \met is corrected for both muon
momentum and the position of the $p\bar{p}$ collision point.  Jets are
found using a fixed-cone algorithm with a cone radius of 0.4 in
$\eta$-$\phi$ space.  To obtain the correct jet energy, this analysis
applies three corrections after jet clustering. We correct for detector
response variations in $\eta$, detector stability, 
and a correction for multiple
interactions in an event. For this analysis, jets are
counted if they have \ET $>$ 15 GeV and $|\eta| < 2.0$ after all the
corrections are applied.
We select events with three or more jets to retain high acceptance for 
$t \bar{t}$ events, allowing one jet to fail our \ET or $\eta$
requirement.

\subsection{b-jet identification}
In this sample, the major background to $t\bar{t}$ is
the electroweak production of a $W$ boson with hadron jets produced by
QCD.  These $W + {\rm jets}$ events usually contain only light 
quark and gluon jets, whereas
signal events always contain two $b$-quark jets.  Thus
identification of $b$-jets ($b$-tag) provides a significant increase in the 
signal-to-background ratio.

We identify $b$-quark jets through the metastable $B$ hadrons in the jet
fragmentation.  
Their $\sim 1$ ps lifetime translates into a
secondary vertex a few millimeters from the primary interaction.  We use the
excellent position resolution of the SVX+ISL to find these secondary
vertices.  The algorithm~\cite{CITE:SECVTX} proceeds as follows: (1)
select at least two good tracks in a jet with both COT and SVX+ISL  
information, (2) search for a high quality secondary
vertex using the selected tracks, (3) measure the distance in the
transverse plane ($L_{T}$) between the primary and the secondary vertices, 
and (4) accept the secondary vertex if
$L_{T}/\sigma(L_{T}) > 3$, where $\sigma(L_{T})$ is the $L_{T}$
resolution. 

Based on simulation of the $b$-tagging algorithm, we determined 
that requiring at least one of the jets in an event 
to be tagged as a $b$-jet is expected to
retain 53$\%$ of the top quark events while removing more than 95$\%$
of the background events~\cite{CITE:SECVTX}.  
The measured cross section depends on
the value of the $b$-tagging efficiency used to extract it.  The
difference between the efficiency in the simulation and that in data
has been measured with a $b$-enriched sample of dijet events in which
an electron is found in one jet (e-jet) and a secondary vertex is
found in the other jet.  From the fraction of e-jets that have an
observed secondary vertex, we find that the simulation has a
$b$-tag efficiency higher than the data by 21$\%$, a factor that is
independent of jet \ET\cite{CITE:SECVTX}.
This difference is corrected for in section VI A and the uncertainty
is discussed in section VI B.

\section{Background Determination}
There are several background sources in the $b$-tagged $l$ +$ \met +
\ge$3 ${\rm jets}$ sample. We give their approximate contributions to
the total background, obtained from both data and theoretical
calculations, although the precise composition is not needed for this
analysis because the spectra are very similar to each other.  The
sources are the production of a $W$ boson accompanied by the QCD
production of heavy flavor quarks ($W$+HF: $\sim40\%$ of the total
number of the background events in $W + \ge 3$ jets),
mis-identification in $b$-jet tagging mainly due to track and vertex
resolution (mistag: $\sim30\%$), a fake $W$ boson associated with a
real or fake lepton (non-$W$: $\sim20\%$), diboson production
($WW/WZ/ZZ$) and electroweak $t\bar{b}$ production (diboson + single
top: $\sim10\%$)~\cite{CITE:SECVTX}.

\subsection{Method}
We find 
that the leading jet \ET is the best discriminant between the 
signal and background among single
jet \ET variables after considering both statistical and systematic 
effects, including the difference between Leading-Order (LO)
and a Next-to-Leading-Order (NLO)
simulation~\cite{CITE:NLO}. Consequently we use the shape of the
leading jet \ET spectrum to
determine the signal and background fractions. 
For the background, we extract from data the
shape for all of the sources except for the small diboson and single
top components.  We assume that the $W$+HF and mistag shapes are the
same, so that we can extract that distribution from the
background-rich sample of events in which no jet is $b$-tagged.  To
avoid subtle kinematic differences based on jet rapidity, we still require 
that at least one jet is taggable, 
i.e., has at least two good tracks that pass
through the SVX+ISL detector. The
assumption that the leading jet \ET in $W$ + multijet events is
independent of the flavor content of the jets is first studied in
Monte Carlo simulations and then tested using events
containing a $W$ boson and either 1 or 2 jets (see section IV
B).

\begin{figure}
\includegraphics[width=12cm]{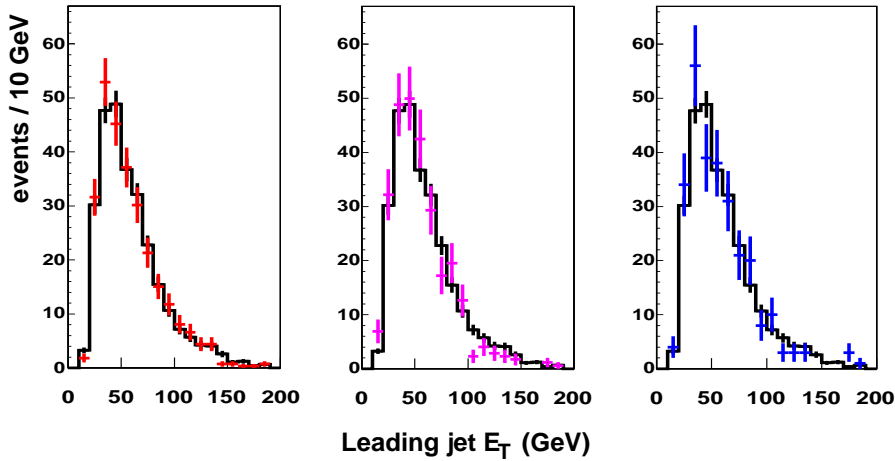}
\caption{ The leading jet \ET spectrum for $W$ plus three light flavor
jets (solid histogram in all plots) compared to $W b\bar{b}$ plus one
light quark (left), $Wc$ plus two light quarks (center), and
$Wc\bar{c}$ plus one light quark (right).  These are ALPGEN+HERWIG
Monte Carlo calculations followed by a simulation of the CDF II
detector. (normalized by area) }
\label{Fig:comp_lfhf}
\end{figure}

\begin{figure}
 \includegraphics[width=12cm]{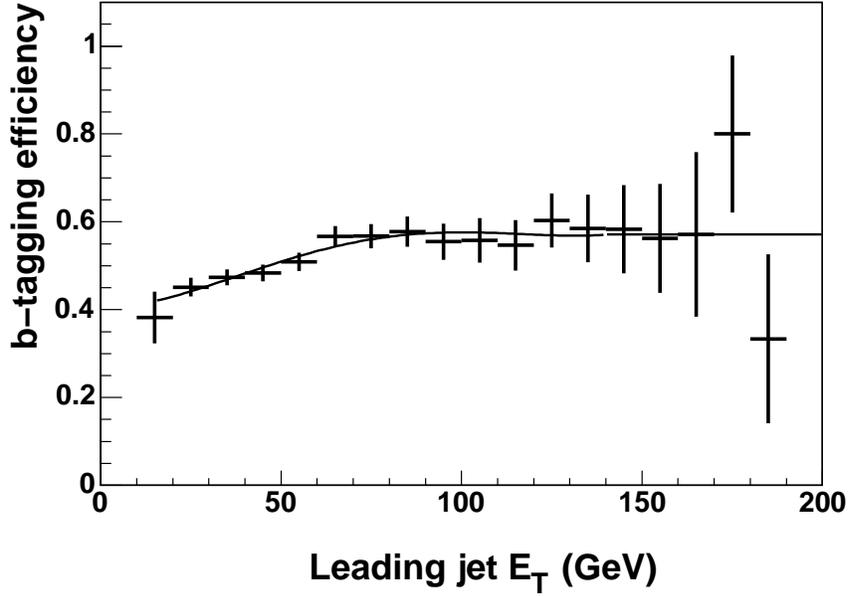}
\caption{ The efficiency of $b$-jet identification as a function of the
leading jet \ET derived from ALPGEN $Wb \bar{b}$ Monte Carlo in the
$W+\ge3$-jet sample.  The curve is a fourth-degree polynomial fit below
140 GeV.  }
\label{Fig:EFFBTAG}
\end{figure}

Figure~\ref{Fig:comp_lfhf} shows the shape comparison of the highest
jet \ET between $W+$ three light flavor jets ($W$+LF) and the various
heavy flavor contributions as predicted by ALPGEN+HERWIG Monte Carlo
calculation followed by a simulation of the CDF II detector.  In
$W$+HF cases, at least one jet is required to be $b$-tagged, while for
$W$+LF, we require that at least one jet is taggable.  We then correct
the $W$+LF shape for the slight \ET dependence in the $b$-tag
efficiency, which is taken from simulation (Fig.~\ref{Fig:EFFBTAG}).

%The leading jet \ET spectrum shape of mistagged events 
%is not {\it a priori} the same as the $W$+HF shape, where the b-jets 
%are tagged, since the shapes may depend 
%on the $b$-tagging algorithm. However, the current algorithm
%provides similar leading jet \ET spectra. Figure~\ref{Fig:mistag} shows
%the $b$-tag rate and the mistag rate as a function of the jet \ET
%obtained from the CDF II jet sample. We apply the mistag rate
%obtained from data to the $W$+LF simulation, and compare it to $W$+HF
%with $b$-tagging requirement for at least one jet in $W$+3 or more
%jets sample (Fig.~\ref{Fig:mistag_HF}).  
%The shapes of mistag and $W$+HF with $b$-tagging requirement
%agree well and therefore the assumption is still valid.  

%\begin{figure}
%\includegraphics[width=12cm]{tagrate.eps} 
%\caption{ The $b$-tag (positive tag in figure) rate and 
%the mistag (negative tag) rate as a function 
%of the jet \ET obtained from the jet data sample~\cite{CITE:SECVTX}. 
%We use a generic 
%jet sample that does not include any significant number of $W$ bosons. 
%The $b$-jet identification algorithm gives  
%similar shapes for the $b$-tag and mistag rates. 
%}
%\label{Fig:mistag}
%\end{figure}

\begin{figure}
\includegraphics[width=12cm]{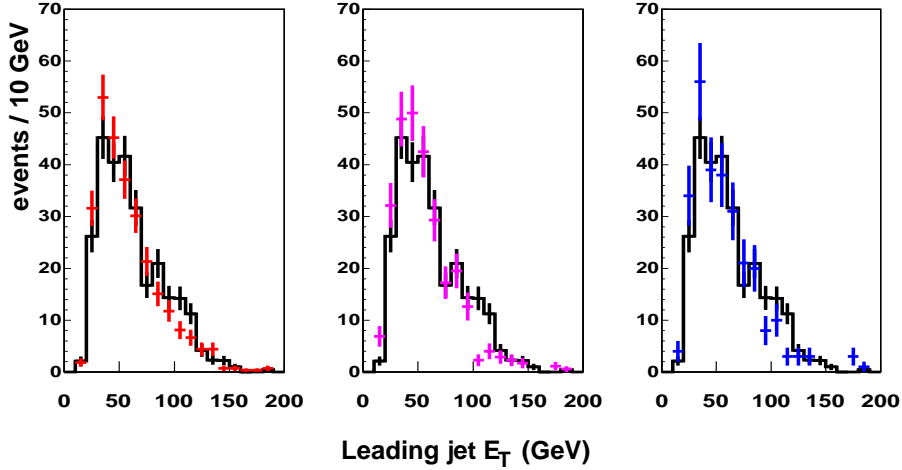} 
\caption{ The leading jet \ET spectrum for mistagged
$W$ plus three light flavor
jets (solid histogram in all plots) compared to $W b\bar{b}$ plus one
light quark (left), $Wc$ plus two light quarks (center), and
$Wc\bar{c}$ plus one light quark (right) in $W$+3 or more jets sample.
For $W$+HF sample, we require at least one tagged $b$-jet.
Note that the mistag
prediction is realistic since it is obtained from jet data.}
\label{Fig:mistag_HF}
\end{figure}

The effect of the $b$-tagging algorithm could be different for $W$+HF,
where a real $b$-jet is tagged, and $W$+LF, where a fake tag is
found.  However the \ET dependences of the tagging efficiency and
mistag probability are similar.  Figure~\ref{Fig:mistag_HF} compares
$W$+HF events with a $b$-tag to $W$+LF after applying the mistag
probability measured in jet data.  The agreement is good.

The other large background comes from events that do not contain a $W$
boson.  The lepton candidate is either misidentified or due to
semi-leptonic decay in a $b$- or $c$-jet.  Such leptons are
typically not isolated in the calorimeter.  Consequently the shape of
the leading jet \ET spectrum for non-$W$ events
is determined from events that still have
large missing transverse energy (\met $>$ 20 GeV), but whose lepton is
not isolated. We assume that the leading
jet \ET shape of this sample is the same as for the non-$W$ background
events in the signal region since lepton isolation is not 
correlated with the \ET of other jets in an event.  The non-$W$
background distribution is added to the other backgrounds with a
relative normalization taken from absolute estimates of the various
background sources.  However, since the non-$W$ jet \ET spectrum is
very similar to those in the $W$+HF and mistag backgrounds, the final
result is insensitive to the non-$W$ fraction.  A systematic
uncertainty is taken based on a large variation in this background
fraction.

The spectra from diboson and single top production are estimated from
ALPGEN+ HERWIG Monte Carlo calculations.  
The leading jet \ET spectrum from single top production is added 
to the total background shape using the
theoretical cross section, which is 6$\%$ of the total background, 
while the diboson
spectra are neglected because these spectra are similar to the 
other dominant background sources, and their contribution is 
expected to be small, $\sim$ 3.0$\%$~\cite{CITE:DB}.

\subsection{Test using $W+1$-jet and $W+2$-jet data}
\begin{figure}
 \includegraphics[width=12cm]{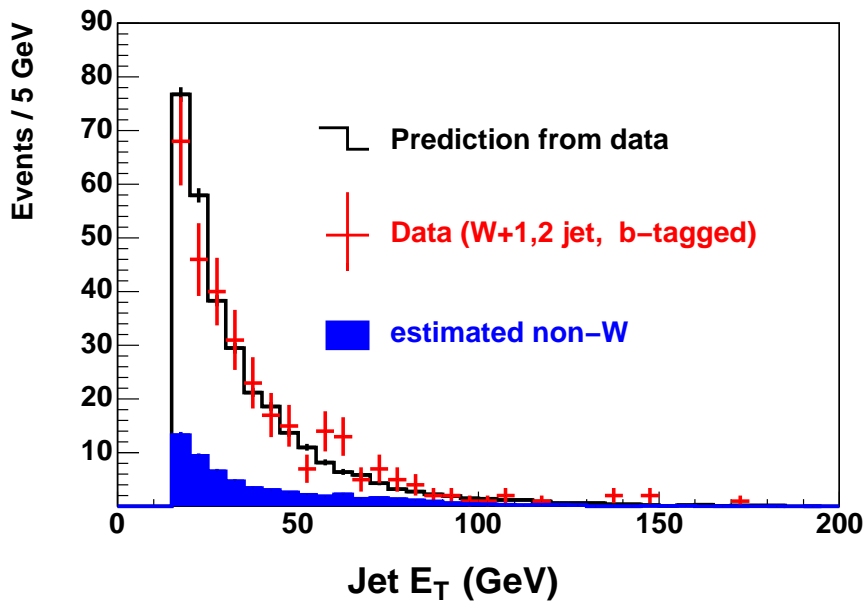} 
\caption{ The
leading jet \ET distribution for the 309 jets in the $W$+1 and 2 jet
data sample.  The open histogram is the total background prediction,
with the dark shaded region the non-$W$ component.  The data points
are the $b$-tagged events with statistical error bars.  }
\label{Fig:w12}
\end{figure}

The performance of the background modeling is tested using events
containing a high \PT lepton, large \metns, and either 1 or 2 jets
(recall the signal sample has 3 or more jets). This sample contains
all of the signal region background sources, but the $t \bar{t}$
contribution is small, only $\sim 4 \%$.  A similar procedure is
applied here as in the $W+\ge3$-jet sample.  The dominant background
shape is taken from events without a $b$-tag, but with at least one
taggable jet, and then a non-$W$ contribution is added with a fraction
($\sim 15.2\%$) determined from absolute background
estimates~\cite{CITE:SECVTX}.  A correction is made for the \ET
dependence of the $b$-tagging efficiency.  The resulting spectrum
should agree well with that of the $b$-tagged events if our method is
valid.  Figure~\ref{Fig:w12} shows the results using 309 jets in the
$W$+1 and 2 jet samples.  The agreement is good, with a
Kolmogorov-Smirnov (KS) test probability of 18$\%$. This value is not
the KS probability itself but the p-value based on 
pseudo-experiments using the maximum difference of the accumulating
distributions.

\subsection{Background shape in the $W+\ge3$-jet sample}
\begin{figure}
 \includegraphics[width=12cm]{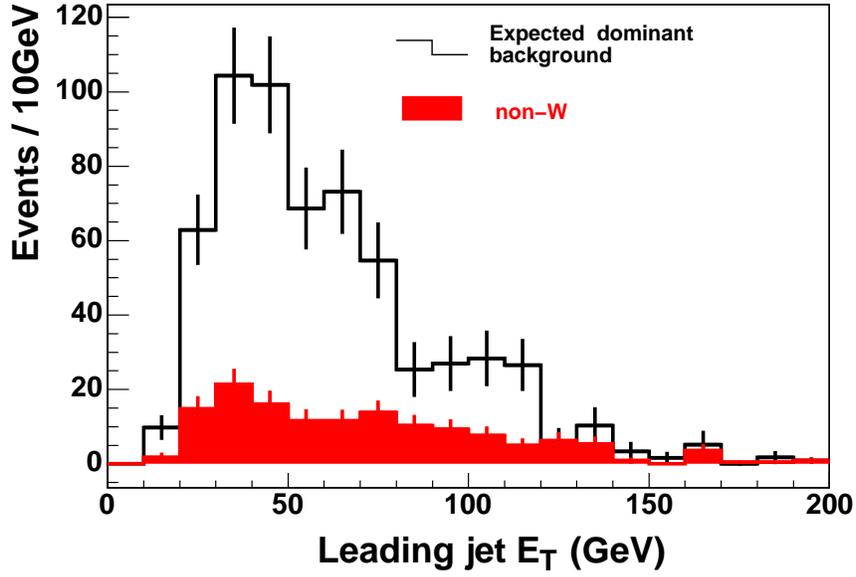} 
\caption{ The calculated dominant background shape 
 ($W$+HF, mistag, non-$W$) in
 $W$+$\ge 3$ jets sample. The shaded part is the non-$W$ portion.  
 Error bars are statistical only.}
\label{Fig:NONW}
\end{figure}

The \ET spectra of the four main backgrounds in the signal sample,
$W+\ge$3 jets, are shown in Fig.~\ref{Fig:NONW}.  The dominant
contribution is taken from the non-$b$-tagged events ($W$+HF, mistag).
These spectra have been corrected for the shape of the b-tag
efficiency, which has been determined from simulation, and is shown in
Fig.~\ref{Fig:EFFBTAG} as a function of the jet \ETns. The small
$t\bar{t}$ contamination, $\sim 6\%$, in the non-$b$-tagged sample is
subtracted iteratively, as described below, so that the amount of this
contamination is consistent with the final $t\bar{t}$ cross section.
The non-$W$ component is $21\%$ of the total and is shown as the
shaded portion.  The small contribution from single-top production is
then added, and the diboson components ($WW/WZ/ZZ$) are neglected
as noted above.

As indicated above, the shape change due to the application of the
$b$-tagging algorithm is applied to the background spectra.  The
efficiency of the algorithm is defined as the number of events that
have at least one $b$-tagged jet divided by the number having at least
one taggable jet.  This function, which drops at very low
\ETns, is fit to a fourth-degree polynomial below 140 GeV and a flat line
above 140 GeV. Possible variations in this shape are considered as a
systematic uncertainty on the cross section measurement.  Note that
only the shape, not the absolute efficiency, is used here.

\begin{figure}
 \includegraphics[width=12cm]{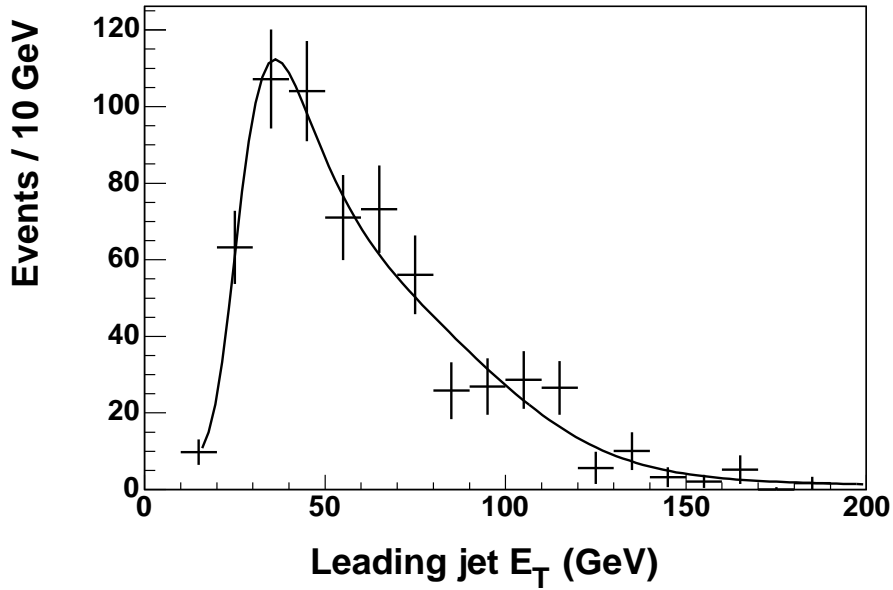}
\caption{ The calculated leading jet \ET distribution for the background
in the $W+\ge3$-jet sample. The fitted curve shows the Landau plus
Gaussian function that is employed in the final unbinned likelihood
fit.  }
\label{Fig:w34_bkg}
\end{figure}

The final background shape is shown in Fig.~\ref{Fig:w34_bkg}.  We fit
this shape for use in the final unbinned log-likelihood fit using a
Landau distribution plus a Gaussian function.  
The fitted parameters
are summarized in Table~\ref{Tab:LG_BKG}, and the fit result is
shown in Fig.~\ref{Fig:w34_bkg}.  The $\chi^{2}/d.o.f.$ for the
agreement between the fit function and the data points is 11.4/10.

\begin{table}
\caption{
  Presented in this table are the fit parameters used to describe 
  the background probability density function shown in Fig.~\ref{Fig:w34_bkg}. 
 (L) and (G) refer to the Landau function and Gaussian parameters, 
  respectively. The variable MPV represents the most probable 
  value of Landau function. The means and sigmas are expressed in GeV.}
  \begin{ruledtabular}
  \begin{tabular}{cc}  
  parameters & values \\ \hline
  height(G) & 30.80$\pm$10.27\\ 
  mean(G) &   66.71$\pm$7.4  \\
  sigma(G) &  32.77$\pm$5.6  \\
  height(L) & 512.8$\pm$79.2 \\
  MPV(L) &    37.0$\pm$3.6   \\
  sigma(L) &  7.67$\pm$1.5   
  \end{tabular}
  \end{ruledtabular}
\label{Tab:LG_BKG}
\end{table}% 

\section{Signal Fraction}
\begin{figure}
 \includegraphics[width=12cm]{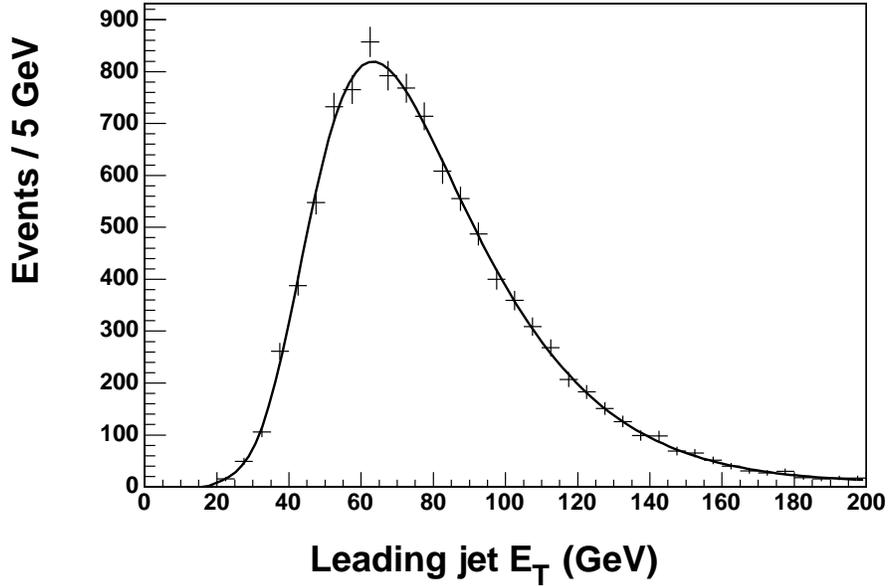} \caption{
The simulated leading jet \ET spectrum for $t\bar{t}$ events for the 
top mass 175 \mgev. 
The smooth curve is a fit to a Landau distribution plus two Gaussians.  }
\label{Fig:TTBAR}
\end{figure}

We use a HERWIG Monte Carlo calculation followed by a full detector
simulation to obtain the $t \bar{t}$ signal shape.
Figure~\ref{Fig:TTBAR} shows the predicted leading jet \ET
distribution for $t \bar{t}$ events.  It is significantly harder than
the background spectrum, making it possible to separate the two
contributions using a fit to the data.  The spectrum in
Fig.~\ref{Fig:TTBAR} is fit to a Landau distribution plus two
Gaussians ($\chi^{2}/d.o.f.$ = 19.9/28). The fitted parameters are
shown in Table~\ref{Tab:LGG_SIG}.  

\begin{table}
\caption{ Fit parameters for the $t \bar{t}$ leading jet \ET
distribution in Fig.~\ref{Fig:TTBAR} using two Gaussians and a Landau 
distribution. (L) and (G) refer to the Landau and Gaussian parameters,
respectively. The variable MPV represents the 
most probable value of Landau function. The means and 
sigmas are expressed in GeV.}
\begin{ruledtabular}
  \begin{tabular}{cc}      
  parameters & values \\ \hline
  height(G1) & 200.8$\pm$26.0 \\
  mean(G1)   &  60.0$\pm$5.0 \\ 
  sigma(G1)  &  35.6$\pm$1.1 \\
  height(G2) & -109.9$\pm$8.3 \\
  mean(G2)   & -76.1$\pm$6.5  \\ 
  sigma(G2)  & 192.9$\pm$14.3 \\
  height(L)  & 3913.5$\pm$192.8 \\
  MPV(L)     & 65.8$\pm$1.1 \\
  sigma(L)   & 13.9$\pm$0.6  
  \end{tabular}
\end{ruledtabular}    
\label{Tab:LGG_SIG}
\end{table}% 

To fit the data to a sum of the signal and background templates, we
use an unbinned likelihood fit with the following form.
\begin{eqnarray}
  {\cal L} &=& \prod^{N}_{i=1} P(E_{Ti};R) \nonumber \\
  &=& \prod^{N}_{i=1} \left[R \:
  P_{\text{signal}}(E_{Ti}) + (1-R) \: P_{\text{background}}(E_{Ti}) 
  \right]\nonumber
\end{eqnarray}
where the signal fraction 
$R$=$\frac{N_{\text{signal}}}{N_{\text{signal}}+N_{\text{background}}}$
is the one free parameter in the fit,  
$P_{\text{signal}}(E_{Ti})$ is the signal probability density
as a function of \ETns, and $P_{\text{background}}(E_{Ti})$ is that of 
the background. We tested 
the ability of this fit procedure to report correct values of the
signal fraction and its uncertainty using a large number of
Monte Carlo pseudo-experiments.  Each pseudo-experiment used the number of
signal and background events in our data sample and used a 
range of signal fractions (R) centered around the value found from our fit
to the data.

As mentioned above, there is a small $t \bar{t}$ contamination in the
untagged data sample that is used to create the background template.
The amount of $t\bar{t}$ that is subtracted when making the template
is determined by an iterative process.  Initially the fit is done
without removing a $t\bar{t}$ component from the background template.
The number of top events reported by the fit is used along with the
$b$-tagging efficiency to calculate the number of top events in the
untagged sample.  A $t\bar{t}$ subtraction in the background template
is then made, and the data are refit. This $t\bar{t}$ contamination   
is determined to be small, $\sim 6\%$, thus only one iteration is necessary.
The final background
template after the iteration is shown in Fig.~\ref{Fig:w34_bkg}.

\begin{figure}
 \includegraphics[width=12cm]{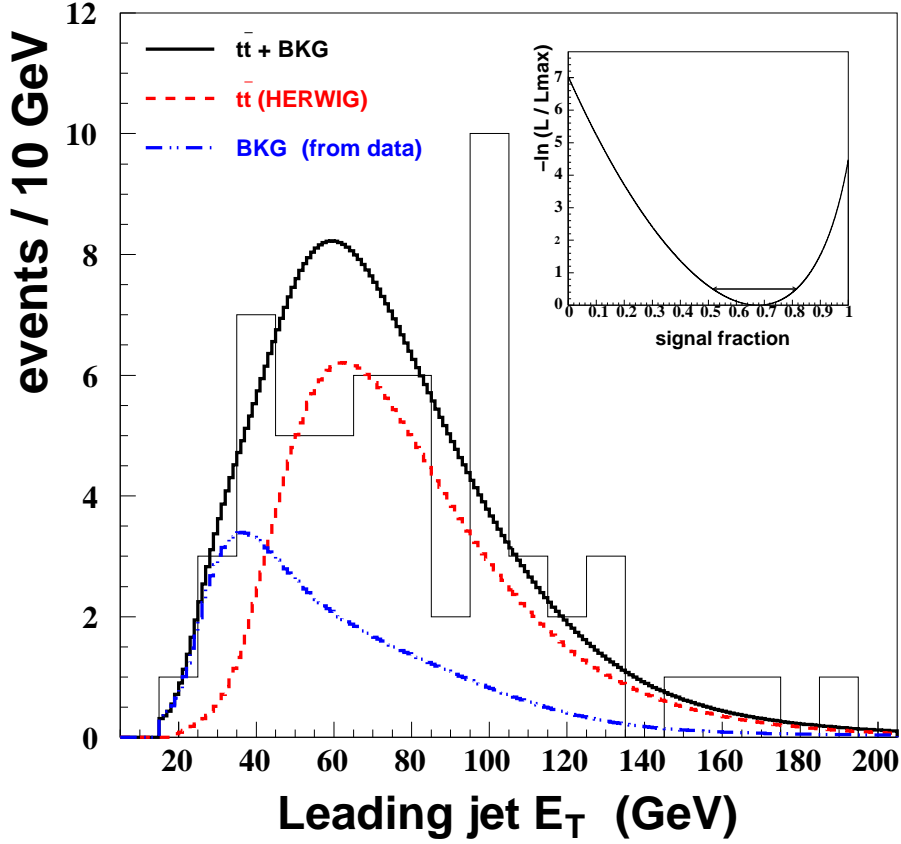} 
\caption {
The fifty-seven candidate events (histogram) with the best fit curve
(solid).  The best fit composition, $t \bar{t}$ (dashed) and
background (dot-dashed), is also shown.  The insert shows $-\ln( {\cal
L}/ {\cal L}_{\text{max}}$) as a function of the signal fraction.  }
\label{Fig:bestfit}
\end{figure}

The result of the fit of the $W+\ge3$-jet data sample is shown in
Fig.~\ref{Fig:bestfit}.  The histogram contains the 57 data events
in which at least one jet has been tagged as a $b$-jet.  The solid
curve is the best fit, with the individual components shown as dashed
($t\bar{t}$) and dot-dashed (background) curves. The insert contains
$-\ln( {\cal L}/ {\cal L}_{\text{max}}$) as a function of signal fraction.
The signal fraction obtained is $R=$0.68$^{+0.14}_{-0.16}$.

\begin{figure}
 \includegraphics[width=12cm]{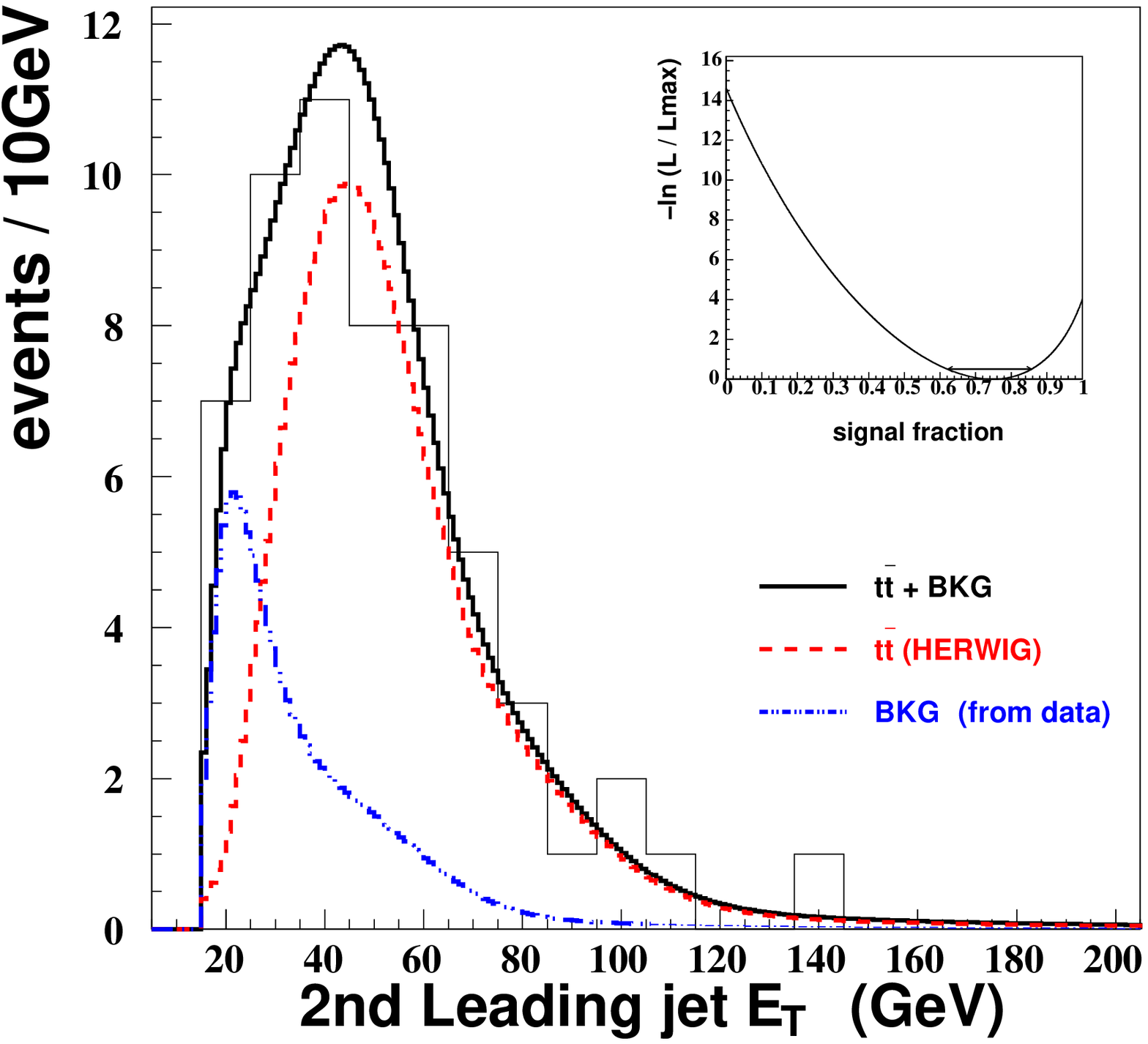} 
\caption {
The fit result using the second leading jet. 
The fifty-seven candidate events (histogram) with the best fit curve
(solid).  The best fit composition, $t \bar{t}$ (dashed) and
background (dot-dashed), is also shown.  The insert shows $-\ln( {\cal
L}/ {\cal L}_{\text{max}}$) as a function of the signal fraction.  }
\label{Fig:bestfit2}
\end{figure}

\begin{figure}
 \includegraphics[width=12cm]{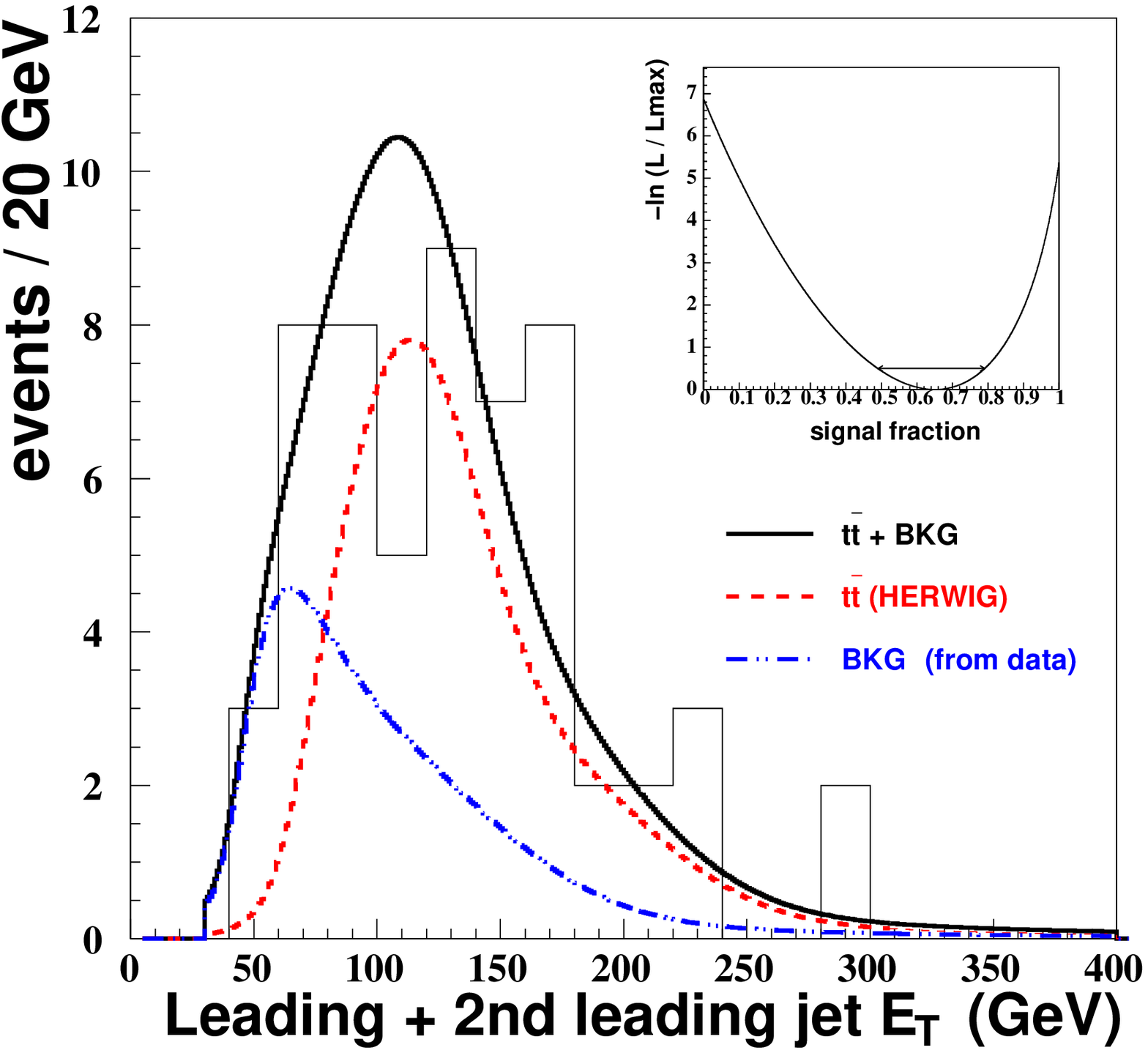} 
\caption {
The fit result using the sum of the leading and the second leading jets. 
The fifty-seven candidate events (histogram) with the best fit curve
(solid).  The best fit composition, $t \bar{t}$ (dashed) and
background (dot-dashed), is also shown.  The insert shows $-\ln( {\cal
L}/ {\cal L}_{\text{max}}$) as a function of the signal fraction.  }
\label{Fig:bestfit12}
\end{figure}

Although we selected the leading jet \ET as the fit variable {\it a
priori}, we have studied other variables to check the robustness of
the result.  For the second leading jet \ET and the sum of the
first and second leading jet \ETns's, we find
% and (c) the scalar sum of the lepton \ET, \met and the \ETns's of 
%the highest three jets, 
signal fractions of $R=$0.75$^{+0.11}_{-0.13}$ (Fig.~\ref{Fig:bestfit2})
and $R=$0.65$^{+0.14}_{-0.16}$ (Fig.~\ref{Fig:bestfit12}),
respectively. The agreement is good.

\section{$t\bar{t}$ cross section}
\subsection{Acceptance and Efficiency}
The $t \bar{t}$ cross section is obtained from the formula, 
\begin{equation}
\sigma (t\bar{t}) = \frac{N_{\text{obs}} \: R_{\text{fit}}}{A_{t \bar{t}}\: 
\epsilon_{t \bar{t}}  \int \! {\cal L} dt}
\end{equation}
where $N_{\text{obs}}$ is the number of candidate $W+\ge3$-jet events with 
at least one $b$-tagged jet (57 events), $R_{\text{fit}}$ is the
signal fraction determined from the likelihood fit
($0.68^{+0.14}_{-0.16}$), and $A_{t\bar{t}}$ is
the geometric acceptance for $t
\bar{t}$ events in the CDF II detector~\cite{CITE:kinematic}.
Note that this acceptance includes the branching ratios.
The parameter $\epsilon_{t \bar{t}}$ is the
detector efficiency for $t\bar{t}$ events~\cite{CITE:SECVTX}, which
includes the trigger, event vertex position, 
event $b$-tagging, and 
the lepton identification efficiencies. 
It also includes the effects due to 
photon conversion, cosmic ray, dilepton, and $Z^{0}$ boson
removal. The quantity $\int {\cal L} dt$ is
the integrated luminosity. The term $A_{t
\bar{t}} \epsilon_{t \bar{t}}$ was determined from a PYTHIA
Monte Carlo~\cite{CITE:PYTHIA} calculation and detector simulation
with a number of individual efficiency components 
determined from the data.  The result for $A\epsilon$
is 4.02$\pm$0.03(stat.)$\pm$0.43(syst.)$\%$. 
The electron (muon) channel contributes $\sim$57$\%$ ($\sim$43$\%$) 
of the total $A\epsilon$.
All calculations have been done using a top quark mass of 
175 \mgev~\cite{CITE:175gev}.  Multiplying
$A\epsilon$ by the integrated luminosity gives the denominator for
equation (1), 6.42$\pm$0.8 pb$^{-1}$. 

\subsection{Systematic uncertainties}
There are a number of sources of systematic uncertainty as summarized in
Table~\ref{Tab:SYSERR}. 
Template shape uncertainties affect the signal fraction determination, while 
other effects mostly impact the acceptance.
Systematic uncertainties in the signal
fraction are determined by a series of pseudo-experiments in which
the generated pseudo-data are changed based on the systematics and
then refit using the original templates.
If the systematic uncertainty affects both the template shapes and the 
acceptance the uncertainty  is taken to be 100$\%$ correlated.

The largest uncertainty originates from the effect of the jet energy
scale on the $t\bar{t}$ simulation.  This comes from a number of
sources including modeling the relative calorimeter response as a
function of $\eta$, the absolute hadron energy scale, the underlying
event contribution, and jet fragmentation~\cite{CITE:kinematic}.  The
largest contributions are due to the $\eta$ correction of the jet energy
and the energy scale uncertainty. The
mean energy of the leading jet from top quark decay is varied by
$\pm$6.1$\%$, or about 5 GeV, and this effect contributes 15.3$\%$ to the final
top cross section uncertainty. The jet energy scale uncertainty does not
contribute to the background template shape systematic uncertainty largely
because it is determined from the data.

There are uncertainties in both the absolute value and \ET dependence
of the $b$-tag efficiencies, which are determined from $b$-jet rich
and generic-jet control samples~\cite{CITE:SECVTX}. The uncertainty of
the absolute $b$-tagging efficiency is dominated by several sources:
statistics of the control data and Monte Carlo sample, composition
uncertainties of the control sample and the branching ratio of the $b$
semi-leptonic decay.  The ratio of the $b$-tag efficiency between the
Monte Carlo and $b$-jet rich control data sample is formed, and we
vary the value within the uncertainties to determine the change in
cross section.  The \ET dependence of the $b$-tag efficiency
uncertainty is determined using the slope difference of two ratios. 
We form as a function of \ET the $b$-tag efficiency ratio from the generic-jet
data and the Monte Carlo simulation (slope and
uncertainty), and also from the $b$-jet rich data sample and Monte Carlo
simulation. A weighted average of the two slopes is used to
determine the overall slope uncertainty, which is then applied to
the Monte Carlo simulation to determine the top quark
cross section uncertainty.

The uncertainty in the background shape due to the statistics of the
background sample is estimated with a series of pseudo-experiments in
which the contents of each bin in Fig.~\ref{Fig:w34_bkg} is varied
independently according to Poisson statistics, with the resulting
distribution refit to get a new background spectrum.  The luminosity
uncertainty comes predominantly from the uncertainty in the total
inelastic cross section.  Uncertainties in the lepton identification
efficiency, which affect the acceptance, are determined from events
using the unbiased tracks of $Z \to ll$ decays in events with multiple
jets~\cite{CITE:kinematic, CITE:SECVTX}.
There are several other efficiency uncertainties due to the trigger
efficiency, the photon conversion veto efficiency, the cosmic ray veto
efficiency, and track finding efficiency. These systematic
uncertainties are summarized in Table~\ref{Tab:SYSERR}. 

The parton distribution functions for (anti-)protons affect not only
the shape of the $t \bar{t}$ signal, but also the acceptances.  These
uncertainties are estimated by varying $\alpha_{s}$ and the parton
distribution functions within the universal fit 
uncertainty.
There are also uncertainties from the amount of gluon radiation in the Monte
Carlo generators.  The amount of initial state radiation is studied
using high mass Drell-Yan dilepton data.  The non-$W$ contribution to
the background shape has an uncertainty both in its relative amount
(to the overall background) and its shape.  The former is estimated
by a $\pm100\%$ variation in the amount of background  measured from
the non-isolated lepton sample.  The shape uncertainty is measured by changing
the nominal mixture of events containing non-isolated electron and muon
candidates ($\sim 3.7:1$ electrons to muons) in the data control sample
to either 100$\%$ electrons or 100$\%$ muons. The shapes are used
in the fit and the change in top quark cross section is reported
as the systematic uncertainty. 

There are shape uncertainties for those spectra obtained from
simulation: $t\bar{t}$ and electroweak $t\bar{b}$ production.  
The uncertainty due to the $t\bar{t}$ shape comes from the
difference between PYTHIA~\cite{CITE:PYTHIA} and
HERWIG~\cite{CITE:HERWIG} simulations.  The theoretical electroweak single top
quark $t\bar{b}$ production cross section uncertainty is small,
which is known to approximately 3$\%$~\cite{CITE:ST}. 
We conservatively
apply an uncertainty of 30$\%$ to the single top cross section. A
background shape uncertainty also results from the uncertainty in the
size of the $t \bar{t}$ contamination in the taggable but untagged
sample and is negligible as we discussed above.  The shape difference
between the mistag and $W$+HF, shown in Fig.~\ref{Fig:mistag_HF}, is
small compared to other systematics.

\begin{table*}
\caption{
 Systematic uncertainties for the $t\bar{t}$ cross section 
are combined assuming different sources are uncorrelated, but
shape and acceptance systematics from each individual source is
100$\%$ correlated.  
}  
\begin{ruledtabular}
  \begin{tabular}{cccc}      
     source  & shape & aceptance & total\\ \hline    
     jet energy scale & $\pm 10.8\%$       & $\pm 4.5\%$  & $\pm 15.3\%$\\ 
     absolute $b$-tag effic. & -----       & $\pm 7.4\%$ & $\pm 7.4\%$\\
     background statistics  & $^{+2.6\%}_{-6.9\%}$ & ----- & $^{+2.6\%}_{-6.9\%}$ \\ 
     luminosity     & -----                & $\pm 5.9\%$ &$\pm 5.9\%$\\ 
     lepton ID      & -----                & $\pm 5.0\%$ & $\pm 5.0\%$\\
    $b$-tag effic. (\ET dependence) & $\pm 1.9\%$ & $\pm 2.5\%$ & $\pm 4.4\%$\\
     parton distribution function & $\pm 3.4\%$ & $\pm 0.8\%$ & $\pm 4.2\%$\\ 
     gluon radiation  & $\pm 0.9\%$        & $\pm 2.6\%$  & $\pm 3.5\%$\\
     non-$W$ (shape)& $\pm 3.0\%$          & ----- & $\pm 3.0\%$\\
     other acceptance syst.  & -----       & $\pm 2.0\%$ & $\pm 2.0\%$\\
     non-$W$ (rate) & $\pm 1.5\%$          & ----- & $\pm 1.5\%$\\     
     $t \bar{t}$ shape  & $\pm 1.5\%$        & ----- & $\pm 1.5\%$\\
%     $t \bar{t}$ in untagged sample  & $\pm 1.5\%$ & ----- & $\pm 1.5 \%$ \\
    $t\bar{b}$ (single top production) & $\pm 0.5\%$   & ----- & $\pm 0.5\%$\\ \hline 
     total          &  $^{+12.4\%}_{-13.9\%}$  & $\pm 12.3\%$ & $^{+20.6\%}_{-21.5\%}$
  \end{tabular}
\end{ruledtabular}    
\label{Tab:SYSERR}
\end{table*}% 

\section{Conclusion}
In conclusion, we have measured the $t \bar{t}$ production cross
section in the lepton plus \met plus jets channel.  $W + \ge 3$ jets 
events were selected with at least one jet identified as a $b$-jet through
secondary vertex reconstruction.  Signal and background were separated
using the shape of the leading jet \ET distribution.  The measured total $t
\bar{t}$ cross section is
6.0$^{+1.5}_{-1.6}$(stat.)$^{+1.2}_{-1.3}$(syst.) pb where we have assumed
a top quark mass of 175 \mgev~\cite{CITE:175gev}. This is
consistent with the Standard Model 
prediction~\cite{CITE:SM} and with the recent result
from CDF in the dilepton channel ($7.0^{+2.4+1.6}_{-2.1-1.1}\pm0.4$
pb)~\cite{CITE:dilepton}. 
The measured cross section depends on the top mass since a heavier top
produces more energetic jets.  This affects both the signal-background
shape discrimination and the acceptance.  A change in the top mass
of $\pm$ 5 GeV/$c^{2}$~\cite{CITE:175gev}
alters the cross section by 6--8$\%$ as shown in Table~\ref{Tab:TOPMASS}.

\begin{table}
\caption{
 The top mass dependence of the measured total $t \bar{t}$ cross section. 
The acceptance and the leading jet \ET 
shape depend on the top quark mass.}  
\begin{ruledtabular}
  \begin{tabular}{cccc} 
     mass (GeV/$c^{2}$) & 170  & 175 & 180 \\ \hline    
     cross section (pb) & 6.4$^{+1.6}_{-1.7}$$^{+1.3}_{-1.4}$ & 6.0$^{+1.5}_{-1.6}$$^{+1.2}_{-1.3}$ & 5.6$^{+1.4}_{-1.5}$$^{+1.1}_{-1.2}$
  \end{tabular}
\end{ruledtabular}    
\label{Tab:TOPMASS}
\end{table}%  

This result also demonstrates that kinematic
determination of the signal fraction using the leading jet \ET
provides good signal to background discrimination. This technique can
be used as an effective constraint in future $t \bar t$ measurements,
such as the top quark mass. This method reduces the sensitivity
to statistical fluctuation of the background because the signal-to-noise
ratio is determined from the top sample itself.

\begin{acknowledgments}
We thank the Fermilab staff and the technical staffs of the
participating institutions for their vital contributions. This work was
supported by the U.S. Department of Energy and National Science
Foundation; the Italian Istituto Nazionale di Fisica Nucleare; the
Ministry of Education, Culture, Sports, Science and Technology of Japan;
the Natural Sciences and Engineering Research Council of Canada; the
National Science Council of the Republic of China; the Swiss National
Science Foundation; the A.P. Sloan Foundation; the Bundesministerium
fuer Bildung und Forschung, Germany; the Korean Science and Engineering
Foundation and the Korean Research Foundation; the Particle Physics and
Astronomy Research Council and the Royal Society, UK; the Russian
Foundation for Basic Research; the Comision Interministerial de Ciencia
y Tecnologia, Spain; and in part by the European Community's Human Potential
Programme under contract HPRN-CT-2002-00292, Probe for New Physics.
%We wish to acknowledge the support of the author community in using
%REV\TeX{}, offering suggestions and encouragement, testing new versions,
\end{acknowledgments}


\end{document}

%% file: author.tex
\affiliation{Institute of Physics, Academia Sinica, Taipei, Taiwan 11529, Republic of China}
\affiliation{Argonne National Laboratory, Argonne, Illinois 60439}
\affiliation{Institut de Fisica d'Altes Energies, Universitat Autonoma de Barcelona, E-08193, Bellaterra (Barcelona), Spain}
\affiliation{Istituto Nazionale di Fisica Nucleare, University of Bologna, I-40127 Bologna, Italy}
\affiliation{Brandeis University, Waltham, Massachusetts 02254}
\affiliation{University of California at Davis, Davis, California  95616}
\affiliation{University of California at Los Angeles, Los Angeles, California  90024}
\affiliation{University of California at San Diego, La Jolla, California  92093}
\affiliation{University of California at Santa Barbara, Santa Barbara, California 93106}
\affiliation{Instituto de Fisica de Cantabria, CSIC-University of Cantabria, 39005 Santander, Spain}
\affiliation{Carnegie Mellon University, Pittsburgh, PA  15213}
\affiliation{Enrico Fermi Institute, University of Chicago, Chicago, Illinois 60637}
\affiliation{Joint Institute for Nuclear Research, RU-141980 Dubna, Russia}
\affiliation{Duke University, Durham, North Carolina  27708}
\affiliation{Fermi National Accelerator Laboratory, Batavia, Illinois 60510}
\affiliation{University of Florida, Gainesville, Florida  32611}
\affiliation{Laboratori Nazionali di Frascati, Istituto Nazionale di Fisica Nucleare, I-00044 Frascati, Italy}
\affiliation{University of Geneva, CH-1211 Geneva 4, Switzerland}
\affiliation{Glasgow University, Glasgow G12 8QQ, United Kingdom}
\affiliation{Harvard University, Cambridge, Massachusetts 02138}
\affiliation{The Helsinki Group: Helsinki Institute of Physics; and Division of High Energy Physics, Department of Physical Sciences, University of Helsinki, FIN-00044, Helsinki, Finland}
\affiliation{Hiroshima University, Higashi-Hiroshima 724, Japan}
\affiliation{University of Illinois, Urbana, Illinois 61801}
\affiliation{The Johns Hopkins University, Baltimore, Maryland 21218}
\affiliation{Institut f\"{u}r Experimentelle Kernphysik, Universit\"{a}t Karlsruhe, 76128 Karlsruhe, Germany}
\affiliation{High Energy Accelerator Research Organization (KEK), Tsukuba, Ibaraki 305, Japan}
\affiliation{Center for High Energy Physics: Kyungpook National University, Taegu 702-701; Seoul National University, Seoul 151-742; and SungKyunKwan University, Suwon 440-746; Korea}
\affiliation{Ernest Orlando Lawrence Berkeley National Laboratory, Berkeley, California 94720}
\affiliation{University of Liverpool, Liverpool L69 7ZE, United Kingdom}
\affiliation{University College London, London WC1E 6BT, United Kingdom}
\affiliation{Massachusetts Institute of Technology, Cambridge, Massachusetts  02139}
\affiliation{Institute of Particle Physics: McGill University, Montr\'{e}al, Canada H3A~2T8; and University of Toronto, Toronto, Canada M5S~1A7}
\affiliation{University of Michigan, Ann Arbor, Michigan 48109}
\affiliation{Michigan State University, East Lansing, Michigan  48824}
\affiliation{Institution for Theoretical and Experimental Physics, ITEP, Moscow 117259, Russia}
\affiliation{University of New Mexico, Albuquerque, New Mexico 87131}
\affiliation{Northwestern University, Evanston, Illinois  60208}
\affiliation{The Ohio State University, Columbus, Ohio  43210}
\affiliation{Okayama University, Okayama 700-8530, Japan}
\affiliation{Osaka City University, Osaka 588, Japan}
\affiliation{University of Oxford, Oxford OX1 3RH, United Kingdom}
\affiliation{University of Padova, Istituto Nazionale di Fisica Nucleare, Sezione di Padova-Trento, I-35131 Padova, Italy}
\affiliation{University of Pennsylvania, Philadelphia, Pennsylvania 19104}
\affiliation{Istituto Nazionale di Fisica Nucleare, University and Scuola Normale Superiore of Pisa, I-56100 Pisa, Italy}
\affiliation{University of Pittsburgh, Pittsburgh, Pennsylvania 15260}
\affiliation{Purdue University, West Lafayette, Indiana 47907}
\affiliation{University of Rochester, Rochester, New York 14627}
\affiliation{The Rockefeller University, New York, New York 10021}
\affiliation{Istituto Nazionale di Fisica Nucleare, Sezione di Roma 1, University di Roma ``La Sapienza," I-00185 Roma, Italy}
\affiliation{Rutgers University, Piscataway, New Jersey 08855}
\affiliation{Texas A\&M University, College Station, Texas 77843}
\affiliation{Texas Tech University, Lubbock, Texas 79409}
\affiliation{Istituto Nazionale di Fisica Nucleare, University of Trieste/\ Udine, Italy}
\affiliation{University of Tsukuba, Tsukuba, Ibaraki 305, Japan}
\affiliation{Tufts University, Medford, Massachusetts 02155}
\affiliation{Waseda University, Tokyo 169, Japan}
\affiliation{Wayne State University, Detroit, Michigan  48201}
\affiliation{University of Wisconsin, Madison, Wisconsin 53706}
\affiliation{Yale University, New Haven, Connecticut 06520}

\author{D.~Acosta}
\affiliation{University of Florida, Gainesville, Florida  32611}
\author{J.~Adelman}
\affiliation{Enrico Fermi Institute, University of Chicago, Chicago, Illinois 60637}
\author{T.~Affolder}
\affiliation{University of California at Santa Barbara, Santa Barbara, California 93106}
\author{T.~Akimoto}
\affiliation{University of Tsukuba, Tsukuba, Ibaraki 305, Japan}
\author{M.G.~Albrow}
\affiliation{Fermi National Accelerator Laboratory, Batavia, Illinois 60510}
\author{D.~Ambrose}
\affiliation{University of Pennsylvania, Philadelphia, Pennsylvania 19104}
\author{S.~Amerio}
\affiliation{University of Padova, Istituto Nazionale di Fisica Nucleare, Sezione di Padova-Trento, I-35131 Padova, Italy}
\author{D.~Amidei}
\affiliation{University of Michigan, Ann Arbor, Michigan 48109}
\author{A.~Anastassov}
\affiliation{Rutgers University, Piscataway, New Jersey 08855}
\author{K.~Anikeev}
\affiliation{Massachusetts Institute of Technology, Cambridge, Massachusetts  02139}
% ...

\collaboration{The CDF Collaboration}
\noaffiliation

%% file: paper.bbl
\begin{thebibliography}{99.}
\bibitem{CITE:plug} We use a cylindrical coordinate system where
$\theta$ is the polar angle to the proton beam from the event 
vertex, $\phi$ is the
azimuthal angle about the beam axis, and pseudorapidity is defined
$\eta = -\ln \tan (\theta/2)$. We define transverse energy as \ET = $E
\sin \theta$ and transverse momentum as $p_{T} = p \sin \theta$
where $E$ is energy measured in the calorimeter and $p$ is
momentum measured by the spectrometer. Missing transverse energy,
\metns, is $-|\Sigma_{i} E^{i}_{T}n_{i}|$, where $n_{i}$ is the unit
vector in the azimuthal plane that points from the beamline to the ith
calorimeter tower.
\end{thebibliography}
